\documentclass[twocolumn,showpacs,showkeys,preprintnumbers,amsmath,amssymb,superscriptaddress]{revtex4}

\usepackage{graphicx}
\usepackage{dcolumn}
\usepackage{bm}

\begin{document}

\title{Diamagnetic Length Scales of Condon Domain Phase in \\ Lifschitz-Kosevich-Shoenberg Approximation}

\author{Nathan Logoboy}

\email{logoboy@phys.huji.ac.il}

\affiliation{Grenoble High Magnetic Field Laboratory, MPI-FKF and
CNRS P.O. 166X, F-38042 Grenoble Cedex 9, France}

\affiliation {The Racah Institute of Physics, The Hebrew University
of Jerusalem, 91904 Jerusalem, Israel}

\affiliation {Institute of Superconductivity, Department of Physics,
Bar-Ilan University, Ramat-Gan 52900, Israel}

\author{Walter Joss}
\affiliation{Grenoble High Magnetic Field Laboratory, MPI-FKF and
CNRS P.O. 166X, F-38042 Grenoble Cedex 9, France}

\affiliation {Universit$\acute{e}$ Joseph Fourier, B.P. 53, F-38041
Grenoble Cedex 9, France}

\date{\today}

\begin{abstract}

Equilibrium properties of non-uniform diamagnetic phase in normal
metals (Condon domains) are studied theoretically in the framework
of Lifschitz-Kosevich-Shoenberg (LKS) approximation. It is found
that characteristic diamagnetic lengths of the phase, e. g. a period
of domain structure and width of interface boundary between domains,
as well as specific surface energy of domain wall, are strongly
affected by electron correlations and depend on temperature,
magnetic field and purity of the sample. The developed theory is in
a good agreement with existent experiment data.

\end{abstract}

\pacs{71.18+y; 75.30.Kz; 75.60.Ch; 75.70.Kw}

\keywords{D. Condon domains; D. Diamagnetic phase transition; D.
dHvA effect}

\maketitle

\section{\label{sec:Introduction}Introduction}

Diamagnetic instability of electron gas in normal metals under
quantizing magnetic field and low temperature is a result of strong
electron correlations induced by magnetic field. It gives rise to a
phase transition with formation of complex domain patterns
\cite{Shoenberg}-\cite{Logoboy1}. The phase transition can occur at
every period of dHvA oscillations and is handled by the tools of
catastrophe theory \cite{Logoboy2}. The symmetric pitchfork
bifurcation gives rise to the second-order phase transition on
temperature at the center of the dHvA period, while the deviation
from the center results in a phase transition of the first order
both in temperature and magnetic field. The diamagnetic phase
transition has received recently much attention due to a number of
unusual phenomena for the physics of diamagnetism, e. g. formation
of complex branch structures \cite{Logoboy1}, \cite{Kramer1}, strong
dependence of magnetic phase diagrams on Fermi-surface topology
\cite{Logoboy3}, \cite{Gordon}, presence of diamagnetic hysteresis
in magnetization curves \cite{Kramer2}, existence of persistent
currents \cite{Logoboy4} which results in a discontinuity of
magnetic induction along the interface boundaries of regular domain
patterns \cite{Condon_Walstedt}, \cite{Kramer1}.

The stratification of the sample into the laminar domain structure,
or Condon domains (CDs), was first observed in a plate-like sample
of silver \cite{Condon_Walstedt} by measuring the NMR frequency
splitting due to a presence of two different kinds of domains.
Later, the increase in absorption of the low-frequency
electromagnetic field (helicons) in aluminum under cooling below
critical temperature was explained by the onset of the diamagnetic
phase transition \cite{Bozhko}. Due to the technical difficulties of
the experimental observation of CDs in normal metals, these results
have remained the only references on the phenomenon of diamagnetic
instability. Recently, the existence of CDs was confirmed by methods
of muon spin-rotation spectroscopy in beryllium, white tin, lead,
aluminum and indium \cite{Solt1}, \cite{Solt2}. The formation of
rather complicated diamagnetic structure in silver was demonstrated
by use of a set micro Hall probes \cite{Kramer1}. Further
development of the experimental technique including the standard ac
method with different modulation levels, frequencies and magnetic
field ramp rates allowed us to reconstruct the magnetization
reversal in beryllium \cite{Kramer2}. The detection of giant
nonlinear response at the crossing critical point, $a \to$1, where
$a=\mu_{0}\max\{\partial M/\partial B\}$ is a differential magnetic
susceptibility, offered a way to construct the diamagnetic phase
diagrams \cite{Kramer1}, \cite{Kramer3}-\cite{Logoboy5}.

Despite numerous experimental evidence for diamagnetic instability
in normal metals, there remain some open fundamental questions
related to the diamagnetic length scales of the CD phase. The
important information about the size of the domains, the domain wall
(DW) width and surface energy of the interface boundaries is still
lacking. The attempt of direct measurement of the period of the
domain structure in plate-like sample of silver by the Hall probe
technique \cite{Kramer1} revealed the value of $\sim$150 $\mu$m
instead of expected one $\sim$30 $\mu$m at the conditions of
experiment (applied magnetic field $\mu_{0}H=$ 10 T, temperature
$T=$ 1.3 K and plate thickness $L \approx$1 mm). Calculation of the
contribution of the short-range interaction (on the scale of
cyclotron radius $r_{c}$) into the free energy density was carried
out by Privorotskii \cite{Privorotskii}. Unfortunately, the
investigation of DW width, specific surface energy of interface
boundary and the size of the domains in \cite{Privorotskii} is
restricted by a limit case $a \to$1. The direct applicability of the
theory of domain structures developed in physics of magnetic
materials \cite{O'Handley} remains under question. So far there is
no reliable theory of the temperature and magnetic field dependence
of the diamagnetic length scales in a full range of existence of the
non-uniform phase.

Motivated by these problems, we present the systematic theoretical
studies of the diamagnetic length scales for normal metals in LKS
approximation. Our investigation is based on the diamagnetic phase
diagrams in a full range of temperature and magnetic field. We
derive equations which allow us to evaluate the DW width $\delta$,
DW-specific surface energy $\sigma$ and the period of the domain
structure $D$ by use of experimentally measured parameter, e. g. the
value of a jump of magnetic induction at the interface boundaries.
We calculate the temperature and magnetic field dependence of
$\delta$, $\sigma$ and $D$ and study the influence of the impurity
of the sample on these characteristics. We show that for a
plate-like sample of silver with thickness $L\sim$1 mm a typical
value of the period of the laminar diamagnetic structure is
$D\sim$0.1 mm which is in two orders of value higher than the
average period of the ferromagnetic domain structure for the sample
of iron of the same shape. Deep inside the diamagnetic phase, the DW
width $\delta$ is almost constant, falling into interval $\sim$1-2
$\mu$m which is of the same order of value as the width of interface
boundary between domains in low-anisotropy magnetic materials such
as thin magnetic films of permalloy (Ni$_{80}$Fe$_{20}$
\cite{O'Handley}). Close to the critical point when
$(r_{c}/L)^{2/3}\ll \alpha \ll$1 ($\alpha =a-$1 is increment of
differential magnetic susceptibility $a$ relative to the critical
value 1), the DW width $\delta$ is scaled as a coherence length
which defines the range of correlations, and diverges with the
critical index $\nu =$1/2 in accordance with mean-field theory.
Approaching at the critical point, a period of the domain structure
decreases, and the domain structure becomes more dense, as a result
of the essential decrease in positive energy of the interface
boundaries. In the nearest vicinity of the critical point when
$\alpha\lesssim (r_{c}/L)^{2/3}$, the stripe domains with
well-defined DWs transform into modulated domain structure (see,
also \cite{Logoboy1}).

The paper is organized as follows. In Sec.~\ref{sec:Model}, we
introduce the model and basic equations. In Sec.~\ref{sec:Results
and Discussions} we calculate the temperature and magnetic field
dependences of DW width, specific surface energy of the DW and
period of the domain structure and discuss the influence of
impurities on these characteristics. Finally, in
Sec.~\ref{sec:Conclusions}, we summarize the conclusions.

\section{\label{sec:Model}Model}

In a single-harmonic approximation the properties of correlated
electrons in normal metals under the conditions of the strong dHvA
effect are described by the free energy functional \cite{Shoenberg}
\begin{equation} \label{eq:Free Energy Functional 1}
G(y;a,x)=a \cos {(x+y)}+\frac{1}{2}y^{2}+\frac{1}{2}a r^{2}_{c}(\partial_{\zeta} y)^{2}, \\
\end{equation}
where the small-scale magnetic field $x=k\mu_{0}(H-H_{a})$ is the
increment of the large-scale internal magnetic field $\mu_{0}H$ and
the applied magnetic field $\mu_{0}H_{a}$, $y=4\pi k M$ is
oscillating part of reduced magnetization, $k=2\pi
F/(\mu_{0}H_{a})^{2}=2\pi /\Delta H$, $F$ is the fundamental
frequency of the dHvA oscillations corresponding to the extremal
cross-section of the Fermi surface, $\Delta H$ is the dHvA period
and $a=\mu_{0}\max\{\partial M/\partial B\}$ is the differential
magnetic susceptibility \cite{Shoenberg}. In physical units $x$ is
of the order of $\thicksim$1-10 mT depending on the properties of
the electron system, while $\mu_{0} H$ is $\thicksim$1-10 T. The
gradient term in Eq.~(\ref{eq:Free Energy Functional 1}) is the
lowest-order term in a full gradient expansion \cite{Privorotskii}
which accounts for the short-range correlations on the scale of
$r_{c}$ ($\zeta$ is coordinate).

In the case of the ellipsoidal Fermi surface, the temperature and
magnetic field dependence of the reduced amplitude of dHvA
oscillations $a$ is defined by \cite{Shoenberg}

\begin{equation} \label{eq:Reduced Amplitude}
a=a_{0}(\mu_{0}H)\frac{\lambda(\mu_{0}H,T)}{\sinh{\lambda(\mu_{0}H,T)}}\exp{[-\lambda(\mu_{0}H,T_{D})]}, \\
\end{equation}
where $\lambda(\mu_{0}H,T)=2\pi^{2}k_{B}T/\hbar \omega_{c}$, $k_{B}$
is the Boltzmann constant, $\hbar$ is the Planck constant,
$\omega_{c}=(e/m_{c})\mu_{0}H$ is cyclotron frequency, $e$ is
absolute value of the electron charge, $m_{c}$ is the cyclotron
mass, and $T_{D}=\hbar /2 \pi k_{B}\tau$ is the Dingle temperature
inversely proportional to the scattering lifetime $\tau$ of
conduction electrons. The limiting amplitude $a_{0}=(H_{m}/H)^{3/2}$
in Eq.~(\ref{eq:Reduced Amplitude}) is the combination of
temperature-independent factors \cite{Shoenberg}, and
$\mu_{0}H_{m}=(10.4\eta\epsilon ^{2}_{F})^{2/3}$ is the maximal
magnetic field above which diamagnetic phase transition does not
occur at any temperature, $\epsilon_{F}$ is Fermi energy in $eV$,
$\eta=m_{c}/m$ and $m$ electron mass. The validity of
Eq.~(\ref{eq:Reduced Amplitude}) is restricted by the application to
the spherical (or almost spherical) Fermi surface sheets, which is
the case of noble metals \cite{Shoenberg}. Equation $a(\mu_{0}H, T,
T_{D})=$1 defines the locus of critical points, e. g. a surface in
three dimensions $\mu_{0}H-T-T_{D}$ which separates the uniform and
CD phases. The calculated phase diagrams are in a good agreement
with the experimental data on measurement of amplitude of the third
harmonic of the $ac$ susceptibility \cite{Kramer3}, justifying the
applicability of $a$ Eq.~(\ref{eq:Reduced Amplitude}) for belly
oscillations in silver.

\begin{figure} [t]
  \includegraphics[width=0.4\textwidth]{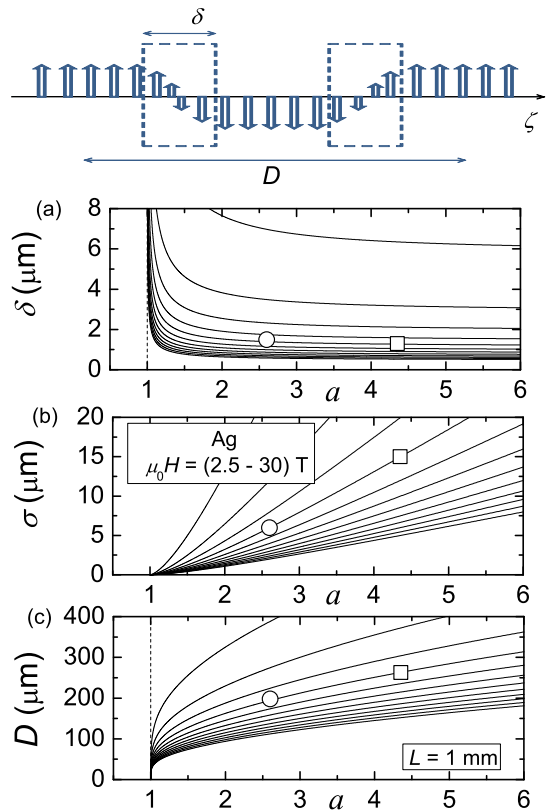}
\caption{ (color online). The upper panel shows a setup of the
system. (a) DW width $\delta=\delta(a)$, (b) specific surface energy
$\sigma=\sigma(a)$, and (c) period of the domain structure $D=D(a)$
are plotted as functions of differential magnetic susceptibility $a$
at different values of the magnetic field $\mu_{0}H$. The magnetic
field increases in steps of 2.5 T starting from 2.5 T (from top to
bottom) which corresponds to decreasing values of $r_{c}=$2.91,
1.45, 0.97, 0.73, 0.58, 0.48, 0.42, 0.36, 0.32, 0.29, 0.26 and 0.24
$\mu$m. Close to the point $a \to $1+0$^{+}$ all characteristics
show critical behavior. The horizontal asymptote for $\delta$ is
2$r_{c}$ (not shown). Circle (square) corresponds to the values of
characteristic lengths calculated at the conditions of experiment
\cite{Condon_Walstedt} (\cite{Kramer1}). The nearest vicinity of the
critical point ($\alpha\lesssim(r_{c}/L)^{2/3}$) where periodic
domain structure transforms into the modulated structure is excluded
from the consideration.} \label{a Characteristics}
\end{figure}

Minimization of the free energy $G$ Eq.~(\ref{eq:Free Energy
Functional 1}) with respect to magnetization $y$ at the center of
dHvA period, $x=0$, leads to differential equation $a \sin {y}-y+a
r^{2}_{c} \partial^{2}_{\zeta \zeta} y=0$ which can be integrated
\begin{equation} \label{eq:Equilibrium}
\int^{y}_{y_{0}} \frac{dy}{f(y;a)}=\frac{\zeta}{a^{1/2}r_{c}},
\qquad
\end{equation}
where $f(y;a)=(y^{2}+2a\cos~y - C)^{1/2}$.
Eq.~(\ref{eq:Equilibrium}) forms the basis for investigation of
non-uniform phases in one-dimensional problems. A proper choice of
integration constant $C$ is dictated by specific boundary
conditions. We assume the existence of periodic domain structure
with alternative magnetization $\pm y_{0}$ in neighboring domains,
defined in explicit form by equation $y_{0}=a\sin~y_{0}$ (see,
Fig.~\ref{a Characteristics}). The period of the domain structure
$D$ is defined by competition between the long-range dipole-dipole
interaction dependent on the size and the shape of the sample, and
short-range electron interaction on the scale of $r_{c}$ which gives
rise to positive energy of interface boundaries. Inserting
$C=y^{2}_{0}+2a\cos y_{0}$ into Eq.~(\ref{eq:Equilibrium}), we
calculate the equilibrium structure of DW with the following
characteristic DW width, $\delta=\delta(a)$, and specific surface
energy of DW, $\sigma=\sigma(a)$ (see, e. g. \cite{Binz})
\begin{eqnarray}
\delta=2a^{1/2}r_{c}\lim_{\epsilon \rightarrow 0} \frac{1}{\ln \epsilon^{-1}}\int^{(1-2\epsilon)y_{0}}_{0} \frac{dy}{f(y;a)}  \label{eq:DWW} \\
\sigma=2a^{1/2}r_{c}\int^{y_{0}}_{0}dy f(y;a). \qquad \label{eq:DW
SSE}
\end{eqnarray}

\begin{figure} [t]
  \includegraphics[width=0.4\textwidth]{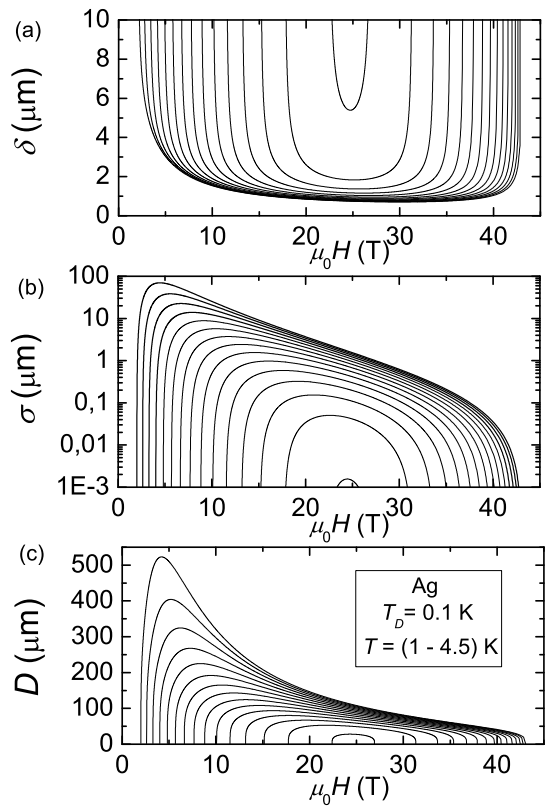}
\caption{ (a) Magnetic field dependence of DW width
$\delta=\delta(\mu_{0}H)$, (b) specific surface energy
$\sigma=\sigma(\mu_{0}H)$, and (c) period of the domain structure
$D=D(\mu_{0}H)$ at Dingle temperature $T_{D}=$0.1 K and different
temperatures $T$. The temperature increases in steps of 0.25 K
starting from 1 K (from bottom to top in (a) and from top to
bottom in (b) and (c)).} \label{H-T}
\end{figure}

\begin{figure}[t]
  \includegraphics[width=0.4\textwidth]{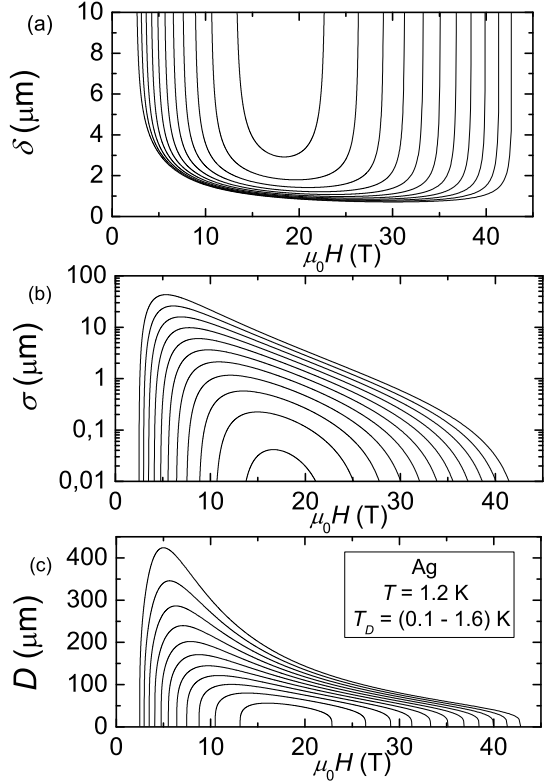}
\caption{ (a) Magnetic field dependence of DW width
$\delta=\delta(\mu_{0}H)$, (b) specific surface energy
$\sigma=\sigma(\mu_{0}H)$, and (c) period of the domain structure
$D=D(\mu_{0}H)$ at $T=$1.2 K and different Dingle temperatures
$T_{D}$. The Dingle temperature increases in steps of 0.15 K
starting from 0.1 K (from bottom to top in (a) and from top to
bottom in (b) and (c)).} \label{H-TD}
\end{figure}

For a plate-like sample of thickness $L$, the standard procedure of
minimization of total free energy of the periodic domain structure
with period $D$, containing two terms, e. g. dipole-dipole energy
$(7/\pi^{3})\zeta(3)y^{2}_{0}D$, where $\zeta(3)$ is zeta-function,
and surface energy of separation of two domains $(2L/D)\sigma$
\cite{Kittel}, allows us to calculate a period of the domain
structure
\begin{equation} \label{eq:P}
D=\frac{(2\pi)^{3/2}}{[7\zeta(3)]^{1/2}}~\frac{(\sigma
L)^{1/2}}{y_{0}}. \qquad
\end{equation}
The DW width $\delta$ Eq.~(\ref{eq:DWW}), specific surface energy of
DW $\sigma$ Eq.~(\ref{eq:DW SSE}) (material-dependent length) and
period of the domain structure $D$ Eq.~(\ref{eq:P}) form a complete
set of characteristic diamagnetic sizes for CD phase. In particular,
the dimensionless characteristic length $\sigma/2L$ (also referred
as material constant) plays an important role in studies of
evolution of the domain structures \cite{Hubert}. The existence of
the well-defined domain structure implies $\delta \ll D/2$.
\begin{figure}[t]
  \includegraphics[width=0.4\textwidth]{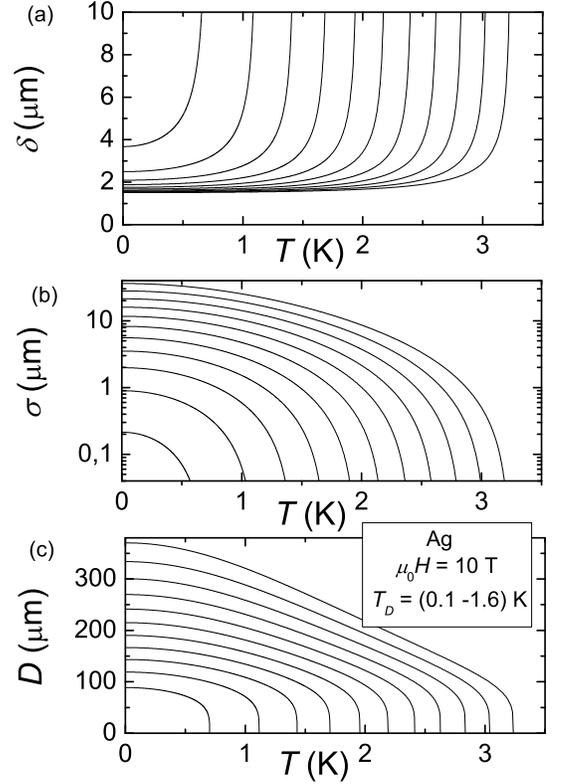}
\caption{ (a) Temperature dependence of DW width
$\delta=\delta(T)$, (b) specific surface energy
$\sigma=\sigma(T)$, and (c) period $D=D(T)$ at $\mu_{0}H=$10 T and
different Dingle temperatures $T_{D}$. The Dingle temperature
increases in steps of 0.15 K starting from 0.1 K (from bottom to
top in (a) and from top to bottom in (b) and (c)).}\label{T-TD}
\end{figure}

\section{\label{sec:Results and Discussions}Results and Discussions}

Close to the critical point when $\alpha \ll $1
Eqs.~(\ref{eq:Equilibrium})-(\ref{eq:P}) can be simplified by using
the expansion of trigonometric functions in powers of $y\le
y_{0}=(6\alpha)^{1/2}\ll$ 1. In this case, the DW structure is
described by the function $y=y_{0}\tanh(\alpha^{1/2}\zeta/r_{c})$
\cite{Logoboy1} with following asymptotic behavior of the
characteristic lengths $\delta$ Eq.~(\ref{eq:DWW}), Eq.~(\ref{eq:DW
SSE}) and $D$ Eq.~(\ref{eq:P}) of the CD phase:
\begin{equation} \label{eq:DWW SSE DSP CP}
\delta = \frac{2r_{c}}{\alpha^{1/2}},~\sigma =
5.7r_{c}\alpha^{3/2},~D =5.3(r_{c}L)^{1/2}\alpha^{1/4}.
\end{equation}
This result is in accordance with the mean-field theory, e. g. near
the critical point the length scale of the fluctuations $\sim
\delta$ has a power law of divergence with the critical index $\nu
$=1/2 and the system has no typical scale length ($\sigma$, $D \to
0$) except of the trivial lower ($r_{c}$) and upper macroscopic (the
size of the system $L$) size scales. It should be noted that
Eqs.~(\ref{eq:DWW SSE DSP CP}) are valid in the range
$(r_{c}/L)^{2/3}\ll\alpha\ll$1 where the low limit value is
evaluated from the condition $\delta\ll D/$2. For $r_{c}\approx$1
$\mu$m and $L=$1 mm we obtain $(r_{c}/L)^{2/3}\approx$0.01. Below
this value (in the nearest vicinity of critical point $a=1$) the
periodic domain structure with well-defined DWs transforms into the
modulated domain structure \cite{Logoboy1}.

Due to the bell-like shape of phase diagrams (see, e. g.
\cite{Gordon}), there is one critical temperature $T_{c}$ at a given
magnetic field and two critical values of the magnetic field
$H_{\pm}$ ($H_{-}<H_{+}$) at a given temperature. Another
possibility for realization of phase transition is related to the
concentration of impurities in the sample, which influence the
amplitude of dHvA oscillations through the scattering lifetime
$\tau$ of conduction electrons. In the vicinity of critical point
$\alpha \to$0$^{+}$ ($T \to T_{c}-$0$^{+}$, or $H \to
H_{\mp}\pm$0$^{+}$, or $T_{D} \to T_{D,c}-$0$^{+}$) the temperature
and magnetic field dependences of $\alpha$ can be represented as
follows
\begin{eqnarray} \label{eq:alpha}
\alpha = \left\{ \begin{array}{ll}
 \lambda_{c}L(\lambda_{c}) t, & t \to 0^{+}, \\
 \lambda_{D,c} t_{D}, & t_{D} \to 0^{+}, \\
 \nu_{\mp} h_{\mp}, & h_{\mp} \to 0^{+},
  \end{array} \right.
\end{eqnarray}
where $L(x)=\coth x -1/x$ is Langevin function and $\pm
\nu_{\mp}=-1.5+\lambda_{\mp}L(\lambda_{\mp})+\lambda^{D}_{\mp}$.
Here, $\lambda_{c}=\lambda(\mu_{0}H,T_{c})$,
$\lambda_{D,c}=\lambda(\mu_{0}H,T_{D,c})$,
$\lambda_{\mp}=\lambda(\mu_{0}H_{\mp},T)$ and
$\lambda^{D}_{\mp}=\lambda(\mu_{0}H_{\mp},T_{D})$. In
Eq.~(\ref{eq:alpha}) $t=$1$-T/T_{c}$, $t_{D}=$1$-T_{D}/T_{D,c}$ and
$h_{\mp}=\pm(H/H_{\mp}-$1) are small increments of temperature,
Dingle temperature and magnetic field for corresponding critical
values $T_{c}$, $T_{D,c}$ and $H_{\mp}$. Substituting
Eq.~(\ref{eq:alpha}) into Eq.~(\ref{eq:DWW SSE DSP CP}), one can
calculate the temperature and magnetic field dependence of $\delta$,
$\sigma$ and $D$ close to the critical point. In particular, if the
phase transition is driven by temperature (at fixed values of
$\mu_{0} H$ and $T_{D}$) we obtain $\delta\sim t^{-1/2}$,
$\sigma\sim t^{3/2}$ and $D\sim t^{1/4}$.

In other limit, $a \to +\infty$, one can neglect the second term in
free energy density Eq.~(\ref{eq:Free Energy Functional 1}). In this
case, the solution of Eq.~(\ref{eq:Equilibrium}) is
$y=2\tan^{-1}\sinh(\zeta/r_{c})$. Thus, we arrive at the following
asymptotic behavior of the diamagnetic length scales \cite{comment}
\begin{equation} \label{eq:Large a}
\delta \approx 2r_{c}, ~\sigma\approx 8ar_{c}, ~D\approx
4.89(r_{c}L)^{1/2}a^{1/2}.
\end{equation}
In the case of $a-1 \gtrsim 1$ one can use the effective free energy
density of interacting electrons
\begin{equation} \label{eq:Free Energy Functional 2}
G=-\frac{1}{2}K\sin^{2}\Theta + \frac{1}{2}A (\partial_{\zeta}
\Theta)^{2},
\end{equation}
where $\Theta=\pi y/2y_{0}$ (see \cite{Logoboy1}, for details).
Parameters $K$ and $A$ are defined as
\begin{equation} \label{eq:Effective Parameters}
K=4a\sin^{2}\frac{y_{0}}{2}(1-a\cos^{2}\frac{y_{0}}{2}), ~
A=a(2r_{c}y_{0}/\pi)^{2}.
\end{equation}
The first term in Eq.~(\ref{eq:Free Energy Functional 2}) is
analogous to the easy-axis crystallographic anisotropy, while the
second one corresponds to the exchange interaction in the physics of
spin magnetism. It confirms the close analogy between easy-axis
anisotropy ferromagnetic sample and the system exhibiting
diamagnetic instability. The structure of DW is well known, it is
described by the equation $y=(2y_{0}/\pi) \tan^{-1}\sinh
(\zeta/\sqrt{A/K})$. In this case, instead of
Eqs.~(\ref{eq:DWW})-(\ref{eq:P}) we obtain the following equations
for the DW width $\delta$, specific surface energy $\sigma$ and
period $D$
\begin{equation}
\delta=\frac{4a}{\pi\psi}r_{c},~\sigma=\frac{4}{\pi}y^{2}_{0}\psi
r_{c},~D=\frac{2^{5/2}\pi(\psi r_{c}L)^{1/2}}{[7\zeta(3)]^{1/2}},
\label{eq:DW W SSE DSP}
\end{equation}
where $\psi=[\sec^{2}(y_{0}/2)-a]^{1/2}$ is a function of $a$. We
note that the form of effective energy density Eq.~(\ref{eq:Free
Energy Functional 2}) captures the essence of exact calculations in
terms of free energy density Eq.~(\ref{eq:Free Energy Functional 1})
even for the regime 0$\le \alpha \lesssim 1$. Thus, in the most
unfavorable case $\alpha \to (r_{c}/L)^{2/3}\ll$1 when the expected
deviation between the results based on Eqs.~(\ref{eq:Free Energy
Functional 1}) and (\ref{eq:Free Energy Functional 2}) is maximal,
the use of effective energy density Eq.~(\ref{eq:Free Energy
Functional 2}) gives the same critical behavior Eq.~(\ref{eq:DWW SSE
DSP CP}) with slightly different numerical factor $\sim$ 0.01.

\begin{figure}[t]
  \includegraphics[width=0.4\textwidth]{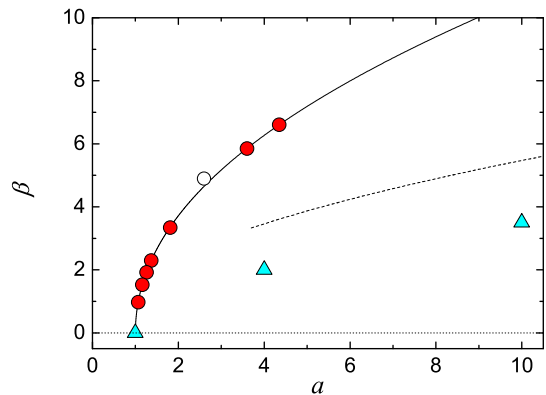}
\caption{ (color online). Parameter $\beta=D/\sqrt{\delta L}$
Eq.~(\ref{eq:beta}) is plotted as a function of differential
magnetic susceptibility $a$ in LKS approximation (solid line). The
dashed line shows the function $\beta=$1.73$a^{1/2}$ evaluated in
simplified dimensional treatment by Shoenberg \cite{Shoenberg}. The
triangles are Condon's calculation of $\beta$ \cite{Condon}. Open
circle corresponds to data \cite{Condon_Walstedt}, close circle are
calculated from the data on measurement of temperature dependence of
the magnetic induction splitting \cite{Kramer1}, as explained in the
text.}\label{beta}
\end{figure}

It follows from Eqs.~(\ref{eq:DW W SSE DSP}) that a monotonic
dependence of the characteristic diamagnetic lengths $\delta,
\sigma$ and $D$ on the applied magnetic field due to dependence of
$r_{c}\sim (\mu_{0}H)^{-1}$ is mediated by strong non-monotonic
dependence on the magnetic field, temperature and purity of a sample
through the differential magnetic susceptibility $a=a(\mu_{0}H, T,
T_{D})$ Eq.~(\ref{eq:Reduced Amplitude}). Both quantities, $r_{c}$
and $a$, can be evaluated directly in experiments on observation of
Condon instability. Fig.~\ref{a Characteristics} shows the
diamagnetic length scales $\delta$, $\sigma$ and $D$ versus
differential magnetic susceptibility $a$ Eq.~(\ref{eq:Reduced
Amplitude}) under various magnetic fields $\mu_{0}H$ in the range
2.5-30 T relevant for appearance of the CD phase in silver
\cite{Kramer3}. At values of $a\in [2, 5]$ typical for experiment
arrangement, the length scales of the CD phase in plate-like sample
of silver are: $\delta\sim$1 $\mu$m, $\sigma\sim$10 $\mu$m and
$D\sim$100 $\mu$m. Under the conditions of experiment on observation
of CD structure by NMR measurement \cite{Condon_Walstedt}
($\mu_{0}H=$ 9 T, $T=$ 1.4 K and $T_{D}=$ 0.8 K) the theory gives
$a=$ 2.6 Eq.~(\ref{eq:Reduced Amplitude}) in accordance with the
value calculated from the splitting of the NMR signal (see, e. g.
\cite{Shoenberg}). It follows from Eq.~(\ref{eq:DW W SSE DSP}) that
$\delta \approx$ 1.5 $\mu$m, $\sigma \approx$ 6 $\mu$m and $D
\approx$ 200 $\mu$m. Similar, in other experiment arrangement, e. g.
$\mu_{0}H=$ 10 T, $T=$ 1.3 K and $T_{D}=$ 0.2 K \cite{Kramer1}, one
can calculate $a=$ 4.35. which gives $\delta \approx$ 1.3 $\mu$m,
$\sigma \approx$ 15 $\mu$m and $D \approx$ 263 $\mu$m.

The results of numerical calculation of the temperature and magnetic
field dependences of the diamagnetic lengths $\delta$, $\sigma$ and
$D$ Eq.~(\ref{eq:DW W SSE DSP}) are illustrated in
Fig.~\ref{H-T}-\ref{T-TD}. In calculation of the period $D$ the
value of $L=$1 mm is used. Fig.~\ref{H-T} shows the magnetic field
dependences of the characteristic lengths of the CD phase at
constant Dingle temperature $T_{D}=$ 0.1 K and different
temperatures $T$. A family of curves demonstrates the existence of
two critical values of the magnetic field in accordance with the
phase diagrams \cite{Gordon}. The functions
$\sigma=\sigma(\mu_{0}H)$ and $D(\mu_{0}H)$ show the existence of
maximums which correspond to the maximum of differential magnetic
susceptibility $a=a(\mu_{0}H)$ with a slight shift into the
low-field range due to the magnetic field dependence of cyclotron
radius $r_{c} \sim (\mu_{0}H)^{-1}$. The growth of temperature
results in the decrease of the interval of values of the magnetic
field where the CD phase exists till it collapses around the value
of $\mu_{0}H \approx$25 T. In Fig.~\ref{H-TD}, the length
characteristics of the CD phase are plotted as a function of the
magnetic field at fixed temperature $T=$1.2 K and different Dingle
temperatures $T_{D}$. The increase in $T_{D}$ due to impurity
scattering leads to the reduction of the amplitude of dHvA
oscillations similar to the temperature effect, but the impurity
effect is more pronounced implying the necessity of using extremely
pure samples in studies of Condon instability. The temperature
dependences of the length characteristics at a fixed value of the
magnetic field $\mu_{0}H=$10 T and different Dingle temperatures are
illustrated in Fig.~\ref{T-TD}. With the increase in the
temperature, the system approaches the point of phase transition
($a\to $1+0$^{+}$) where $\delta$, $\sigma$ and $D$ show critical
behavior in a accordance with Eqs.~(\ref{eq:DWW SSE DSP CP}) and
(\ref{eq:alpha}).

\begin{figure}[t]
  \includegraphics[width=0.4\textwidth]{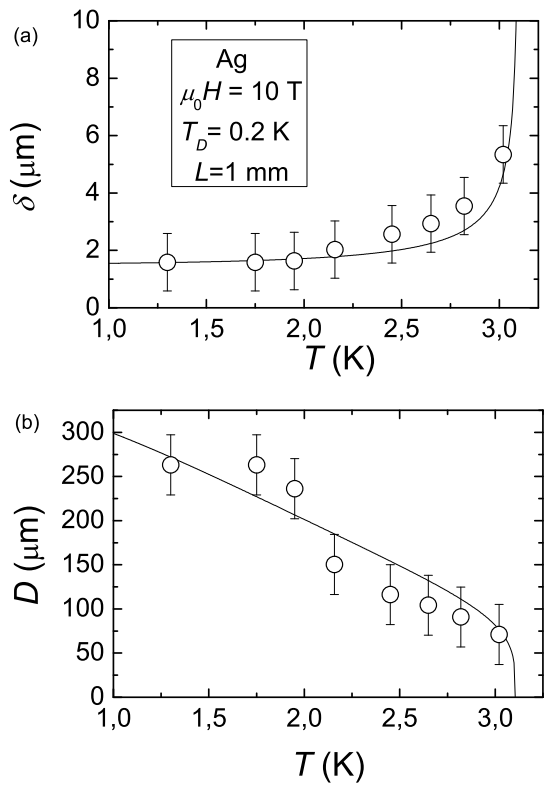}
\caption{ (color online). Temperature dependence of the DW width
$\delta$ (a) and period of the domain structure $D$ (b) at the
conditions of the experiment \cite{Kramer1}. The solid lines
correspond to the theory, the circles are calculated from the
temperature dependence of the measured jump of magnetic induction at
the interface boundaries \cite{Kramer1}.}\label{T_Kramer}
\end{figure}
It is convenient to introduce a parameter $\beta=D/(\delta L)^{1/2}$
independent of the width of the plate $L$ and commonly used in
studies of domain structures. It follows from Eq.~(\ref{eq:DW W SSE
DSP}) that $\beta=\beta(a)$ is defined entirely by the properties of
correlated electron gas through differential magnetic susceptibility
$a$ Eq.~(\ref{eq:Reduced Amplitude})
\begin{equation}
\beta=\frac{(2\pi)^{3/2}}{[7\zeta(3)]^{1/2}}\frac{\psi(a)}{a^{1/2}}.
\label{eq:beta}
\end{equation}
In Fig.~\ref{beta} parameter $\beta$ Eq.~(\ref{eq:beta}) is plotted
as a function of $a$ in LKS approximation together with the
corresponding estimates due to Shoenberg \cite{Shoenberg} and Condon
\cite{Condon}. Fig.~\ref{beta} illustrates essential discrepancy
between the theoretical results. In analysis of the domain structure
by Shoenberg \cite{Shoenberg}, there are two assumptions. The first
assumption is related to the amplitude of dHvA oscillations, e. g.
the only case of extremely large values of $a \to \infty$ when
$y_{0}= \pi$ was considered. The second assumption involves the
energy of interface boundary $\sigma$ and plays a crucial role in
analyzing the period of the domain structure. This energy was
evaluated roughly in the {\it order of value} as $\sigma \sim a
\delta$. It gives the correct asymptotic behavior $D \sim a^{1/2}$,
but the important numerical factor is missing (see,
Eq.~(\ref{eq:Large a})). As a result, the minimization of the total
energy leads to parameter $\beta=$ 1.73 $a^{1/2}$. Undoubtedly, the
coefficient in the expression remains under question which was also
marked by Shoenberg \cite{Shoenberg} who supposed to use Condon's
results instead. Unfortunately, we cannot discuss Condon's
estimations \cite{Condon} which are represented in \cite{Shoenberg}
without prove. As we see below, Condon's result is in contradiction
with experiment \cite{Kramer1}. In order to calculate a quantity
such as the surface energy of the DW Eq.~(\ref{eq:DW SSE}), it is
necessary to have an expression for the DW structure
(\ref{eq:Equilibrium}). The calculations in LKS approximation allow
us to obtain a correct expression for $\beta$ offering the missing
numerical factors for $\sigma$ and $\beta$. In particular, the use
of Eq.~(\ref{eq:Large a}) which is true in a limit $a \to \infty$
results in $\sigma=$ 4 $\delta a^{1/2}$ and $\beta=$ 3.45 $a^{1/2}$.

The confirmation of the validity of our studies comes from the
analysis of the data obtained by the Hall probe technique
\cite{Kramer1}. A set of micro Hall probes was used for detection of
local induction at the surface of the pure silver single crystal
2.4$\times$1.6$\times$1.0 mm$^{3}$ ($T_{D}=$ 0.2 K) in the magnetic
field up to 10 T and temperature interval $T=$1.3$-$3 K. The
detection of the inhomogeneous induction was attributed to the
presence of the CD structure with the period evaluated as being
certainly larger than the distance of $\approx$ 150 $\mu$m between
the edge probes (under limit for period). This result is in
contradiction with the proposed in \cite{Kramer1} expression
$D\approx (L\delta)^{1/2}$ ($\beta=$ 1) that gives much lower value
for $D\approx$ 30 $\mu$m. Under the conditions of experiment, $a=$
4.35 Eq.~(\ref{eq:Reduced Amplitude}), $r_{c}=$ 0.7 $\mu$m. Thus,
according to Shoenberg \cite{Shoenberg}, $\beta=$ 1.73
$a^{1/2}\approx$ 3.61 and $D\approx$ 140 $\mu$m which is close to,
but still less than experimentally determined under limit 150
$\mu$m. The estimates due to Condon results in the lower value of $D
\approx$ 77 $\mu$m ($\beta \approx$ 2 \cite{Condon}). On the other
hand, calculations in the framework of the presented theory give the
same value of $\delta \approx$ 2 $r_{c} \approx$ 1.4 $\mu$m, but the
larger value of $D=$ 263 $\mu$m which is more reasonable because it
is above the detected lower limit.

In the case of $\alpha \gtrsim$1 which is typical for the experiment
arrangement on investigation of the CD phase, the change of the
magnetic field within the period of dHvA oscillations does not
affect the average jump of magnetic induction at the interface
boundaries of domain patterns, $\delta B=2y_{0}/k$ \cite{Shoenberg},
defined by the local magnetization at the center of the dHvA period,
$y_{0}=y_{0}(a)$. It offers a way to calculate the expected values
of diamagnetic length scales by means of measurement of the value of
$\delta B$. The value of differential magnetic susceptibility $a$
deduced from the measured jump of magnetic induction can be used for
evaluation of parameter $\beta$ Eq.~(\ref{eq:beta}) and diamagnetic
length scales Eq.~(\ref{eq:DW W SSE DSP}). The results of numerical
calculation of $\beta$ from the measurements of temperature
dependence of the magnetic field distribution \cite{Kramer1} are
shown in Fig.~\ref{beta}. Fig.~\ref{T_Kramer} illustrates the
temperature dependence of the DW width $\delta$ and period of the
domain structure $D$ under the conditions of the experiment
\cite{Kramer1}. There is a good agreement between the theory and the
data.

\section{\label{sec:Conclusions}Conclusions}

Characteristic diamagnetic length scales of the Condon domain phase
in normal metals under quantizing magnetic field and low temperature
are studied theoretically in LKS formalism. The results of
calculation show that temperature, magnetic field and purity of the
sample affect greatly the width of domains and the specific surface
energy of DWs, but have little influence on the width of interface
boundaries if the system is far enough from the critical point. Well
inside the diamagnetic phase, the DW width falls into interval
$\sim$1-2 $\mu$m which is of the same order of value as width of the
interface boundary between domains in low-anisotropy magnetic
materials such as permalloy thin magnetic films. For a plate-like
sample of silver with thickness $\sim$1 mm a typical value of the
period of the diamagnetic domain structure is $\sim$0.1 mm which is
in two orders of value higher than the average period of
ferromagnetic domains for the sample of iron of the same shape.

Approaching at the critical point when
$(r_{c}/L)^{2/3}\ll\alpha\ll$1, the DW width being a characteristic
of the range of correlations (coherence length) diverges $\delta
\sim \alpha^{-1/2}$ with the critical index $\nu $=1/2 in accordance
with mean-field theory, while the period of the domain structure
goes to zero, e. g. the domain structure becomes more dense. The
effect of "shrinking" of the domain structure is a result of
essential decrease in positive energy of the interface boundary,
$\sigma$, which scales as $\sim y_{0}^{3}\sim \alpha^{3/2}$. The
energy of long-range interaction, $G_{dd}$, scales as $\sim
y^{2}_{0}$. Thus, $D\sim \sigma^{1/2}/y_{0}\sim \alpha^{1/4}$ (see,
Eq.~(\ref{eq:DWW SSE DSP CP})). We show that diamagnetic length
scales can be calculated from the data on measurement of the jump of
magnetic induction at the interface boundaries of domain patterns.

Theoretical results are in agreement with available experimental
data. We hope that our studies will stimulate further experimental
investigation of diamagnetic length scales in normal metals at the
conditions of strong dHvA effect.

\begin{acknowledgments}

We are indebted to V. Egorov, I. Sheikin and D. Golosov for
fruitful discussions.

\end{acknowledgments}


\begin{thebibliography}{99}
\bibitem{comment} By direct calculation \cite{Privorotskii} one can show that the next non-zero term in gradient expantion of free energy Eq.~(\ref{eq:Free Energy Functional 1}) is $- (5/48) a
r^{4}_{c}(\partial^{2}_{\zeta\zeta})^{2}$. For $a \to +\infty$
($\delta \to 2r_{c}$), the additional term is a factor 10 to 100
smaller than the lowest gradient term and can be taken into account
by itteration resulting in a change of DW width withing $\approx$10
$\%$. For intermidiate values of $a$ ($\delta>>d_{c}=2r_{c}$) the
error should be smaller due to the presence of prefactor
$(d_{c}/\delta)^{4}<<1$.
\end{thebibliography}
\end{document}